\documentstyle[12pt]{article}

\title{Weak Decoherence and Quantum Trajectory Graphs} 
\author{Eric Chisolm\thanks{echisolm@physics.utexas.edu} \ and E. C. G. 
        Sudarshan \\ Center for Particle Physics, Physics Department \\ 
        University of Texas at Austin \\ Austin, TX  78712 \and Thomas F. 
        Jordan \\ Physics Department, University of Minnesota \\ Duluth, MN  
        55812}

\input{psfig}

\begin{document}

\maketitle

\begin{abstract}
Griffiths' ``quantum trajectories'' formalism is extended to describe weak 
decoherence.  The decoherence conditions are shown to severely limit the 
complexity of histories composed of fine-grained events.
\end{abstract}

\vspace{-5.75in} 
\begin{flushright} CPP-94-33 \end{flushright}
\vspace{5.6in}

In response to the increasingly popular opinion that the Copenhagen 
interpretation of quantum mechanics raises more questions than it answers 
\cite{om2} and a desire to treat the entire universe quantum mechanically, 
Gell-Mann and Hartle \cite{SFI,misner,D47} have worked to create an alternative
interpretation of quantum theory, expanding upon earlier work by Griffiths 
\cite{consis} and Omn\`{e}s \cite{om1}.  Their scheme emphasizes not individual
events but Griffiths' notion of a {\em history}, a sequence of events at a 
succession of times, and they assert that the histories to which one assigns 
probabilities are distinguished not by measurements made by an external 
classical ``observer'' but by the extent to which they satisfy certain
``consistency'' or ``decoherence'' conditions guaranteeing compliance with the
classical rules of probability.  

Yet basic questions remain largely unanswered:  How restrictive are the 
decoherence conditions?  What kinds of histories decohere?  Do they occur in 
sufficient variety to describe the physical world?

These questions have led us to investigate several aspects of decoherence.  We 
have extended Griffiths' ``quantum trajectories'' formalism \cite{graph} to 
describe {\em weakly} decohering sets of histories.  We have found severe 
limits on the structure of fine-grained decohering histories. 

Following Gell-Mann and Hartle, we let an {\em event} be described by a
projection operator $P_{\alpha}$.  If $P_{\alpha}$ is one-dimensional we say 
the event is {\em fine-grained}; otherwise the event is {\em coarse-grained}.  
A {\em complete} set of events $\{ P_{\alpha} \}$ forms a resolution of the 
identity: 
\begin{equation}
  \sum_{\alpha} P_{\alpha} = I \ \ {\rm and} \ \ P_{\alpha} P_{\beta} = 
  \delta_{\alpha\beta} P_{\beta}. 
\end{equation}
Let $\{P_{\alpha_{k}}(t_{k})\}$ be the complete set of events (in the 
Heisenberg picture) at time $t_{k}$. The probability that event $P_{\alpha_{1}}
(t_{1})$ will occur at time $t_{1}$, $P_{\alpha_{2}}(t_{2})$ at time $t_{2}$, 
\ldots, and $P_{\alpha_{n}}(t_{n})$ at time $t_{n}$ is \cite{misner}
\begin{eqnarray}
 \lefteqn{ p(\alpha_{1}, \alpha_{2}, \ldots, \alpha_{n}) = } \nonumber \\
 & & {\rm Tr\,}(P_{\alpha_{n}}(t_{n}) \ldots P_{\alpha_{2}}(t_{2}) 
                P_{\alpha_{1}}(t_{1}) \rho P_{\alpha_{1}}(t_{1}) 
                P_{\alpha_{2}}(t_{2}) \ldots P_{\alpha_{n}}(t_{n}))
 \label{projprop}
\end{eqnarray}
for an initial state described by a density operator $\rho$.  With this 
sequence of events we associate the {\em history} $C_{\alpha}$ defined by 
\begin{equation}
   C_{\alpha} = P_{\alpha_{n}}(t_{n}) \ldots P_{\alpha_{2}}(t_{2}) 
                P_{\alpha_{1}}(t_{1}), 
\end{equation}
in terms of which (\ref{projprop}) becomes
\begin{equation}
   p(\alpha) = {\rm Tr\,}(C_{\alpha} \rho \mbox{$C_{\alpha}$}^{\dag}). 
   \label{prob} 
\end{equation}
With this expression in mind, we define the {\em decoherence functional} 
between histories $C_{\alpha}$ and $C_{\beta}$ for an initial state $\rho$ by
\begin{equation} 
  D(\alpha, \beta) = {\rm Tr\,}(C_{\alpha} \rho \mbox{$C_{\beta}$}^{\dag}),
\end{equation}
and we say $\{C_{\alpha}\}$ forms a {\em weakly decohering} set of histories 
iff 
\begin{equation} 
  {\rm Re\,}D(\alpha, \beta) = 0 \ \ {\rm for\ all} \ \ \alpha \neq 
            \beta. \label{decohere1} 
\end{equation}
This condition guarantees that the probabilities associated with the histories 
in $\{C_{\alpha}\}$ obey the classical rules of probability \cite{misner}.  
For such a set of histories, the decoherence condition (\ref{decohere1}) and
the probability formula (\ref{prob}) may be combined in the equation
\begin{equation} 
  {\rm Re\,}D(\alpha, \beta) = p(\alpha)\delta_{\alpha\beta}. 
     \label{decohere2} 
\end{equation}
A set of histories which satisfies the stronger condition
\begin{equation} 
  D(\alpha, \beta) = p(\alpha)\delta_{\alpha\beta} \label{stdec} 
\end{equation}
is said to exhibit {\em medium decoherence}.  This condition is sufficient but 
not necessary to ensure compliance with the classical rules of probability.  We
will consider both kinds of decoherence.  Finally, we will speak of individual 
histories $C_{\alpha}$ and $C_{\beta}$ decohering if they satisfy 
(\ref{decohere2}) (or (\ref{stdec})).

Histories with different final events always decohere and any history which 
occurs with zero probability decoheres with all other histories.  Further, any 
decohering set of histories can be extended by inserting between any two times 
a set of events identical (in the Heisenberg picture) to those at the earlier 
or later time; the set of histories that results still decoheres.  This 
corresponds to inserting a set of events in the Schrodinger picture which 
matches the earlier or later set aside from unitary evolution to the new time. 
We call this a {\em congruent extension}, since the new events are congruent 
with the old ones.  In light of this, we will look for sets of histories with 
more than one nonzero-probability history but without congruent extensions.

\begin{figure}[p]
  \centerline{\psfig{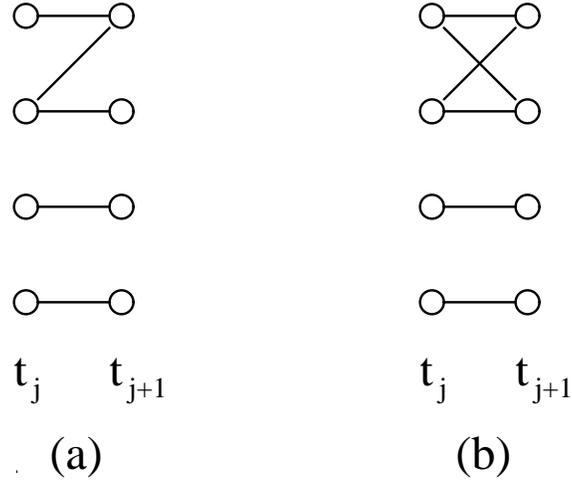}}
  \caption{Griffiths trajectory graphs.  (a) A candidate for a nontrivial 
           transition in which the number of connected events does not 
           increase.  This graph is forbidden by the orthogonality of different
           events at $t_{j}$.  (b) Another candidate for the same transition.  
           The orthogonality of different events at $t_{j}$ demands that at 
           least two events at $t_{j+1}$ be doubly connected.} 
  \label{unfor}
\end{figure}

\begin{figure}[p]
  \centerline{\psfig{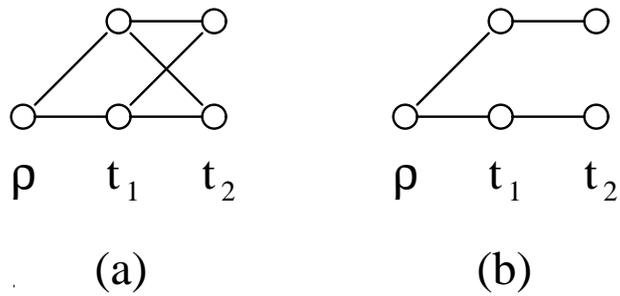}}
  \caption{Trajectory graphs with a specified initial state.  (a) A graph 
           corresponding to a weakly decohering set of histories of a two-level
           system.  (b) A graph corresponding to a set of histories exhibiting 
           medium decoherence.}
  \label{twolev}
\end{figure}

\begin{figure}[p]
  \centerline{\psfig{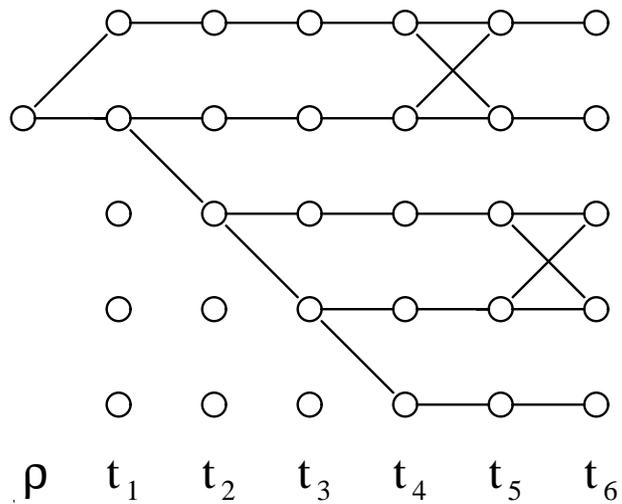}}
  \caption{A set of decohering histories in a five-dimensional space with 
           exactly \mbox{$5 + [\frac{5}{2}] - 2 = 5$} noncongruent 
           transitions.}
  \label{maxnon}
\end{figure}

We will use Griffiths' graphical representation \cite{graph} of ``consistent 
histories,''  in which he represents the set of possible events at each time 
with an orthonormal basis of the system's state space.  (In this formalism, 
every event is fine-grained.)  He represents the set of histories produced by 
this choice of events with a {\em trajectory graph}, in which each event at 
time $t_{j}$ corresponds to a node in the $j^{th}$ column of the graph and a 
line is drawn between nodes in adjacent columns iff the transition amplitude 
between the corresponding events is nonzero.  Figure~\ref{unfor} presents two 
examples of such graphs.  Each path (unbroken line through two or more nodes) 
through a trajectory graph represents a nonzero-probability history with 
initial state given by the first node in the path.  The set of histories 
described by the graph satisfies the {\em noninterference condition} if any two
nodes are connected by at most one path.  We will show immediately below that 
the noninterference condition is equivalent to {\em medium} decoherence of the 
set of histories with any node in the graph as the initial state.  However, we 
wish to follow in the spirit of Gell-Mann and Hartle, in which decoherence is a
function of the initial state as well as the histories themselves.  Further, 
since medium decoherence is a more stringent requirement than is actually 
necessary, we would like to have a condition for {\em weak} decoherence in 
terms of these graphs.  As we will also prove below, the required condition is 
that {\em at most {\rm two} distinct paths connect any two events, and if there
are two paths, the phases of the corresponding amplitudes differ by 
$\frac{\pi}{2}$}.  Thus, we will use a modified form of Griffiths' quantum 
trajectory formalism in which (1) we specify the initial state (producing what 
Griffiths would call an {\em elementary family} of trajectories) and (2) we 
impose the requirement of weak, not medium, decoherence.  Figures~\ref{twolev} 
and \ref{maxnon} provide examples of such graphs.

Both decoherence conditions mentioned above are special cases of the following 
theorem. \vspace{.2in} \\
{\bf Theorem 1.}  Suppose $\{C_{\alpha}\}$ is a decohering set of histories 
     with initial state $\rho$ and suppose that two possible events $|j \rangle
     \langle j|$ at time $t_{j}$ and $|k \rangle \langle k|$ at a later time 
     $t_{k}$ are fine-grained.  If at least one history leading to event $|j 
     \rangle \langle j|$ occurs with nonzero probability, then of all histories
     which lead from $|j \rangle \langle j|$ to $|k \rangle \langle k|$, at 
     most {\em two} occur with nonzero probability.  If two occur, then the 
     phases of the corresponding amplitudes differ by $\frac{\pi}{2}$.  If the 
     set $\{C_{\alpha}\}$ exhibits {\em medium} decoherence, at most {\em one} 
     history leading from $|j \rangle \langle j|$ to $|k \rangle \langle k|$ 
     occurs with nonzero probability. 
\vspace{.1in} \\
{\bf Proof.}  Any history leading from $| j \rangle \langle j |$ to $|k \rangle
\langle k|$ can be written as
\begin{equation} 
  C_{\alpha} = |k \rangle \langle k| D_{\alpha}|j \rangle \langle j| 
     \label{abc} 
\end{equation} 
and the decoherence condition (\ref{decohere1}) applied to any two histories
which include $C_{\alpha}$ and $C_{\beta}$ where $\alpha \neq \beta$ can be 
reduced to
\begin{equation} 
  {\rm Re\,} \langle k | D_{\alpha} | j \rangle \langle k | D_{\beta} | j 
         \rangle^{*} = 0. \label{orth} 
\end{equation}
If both amplitudes are nonvanishing, then 
\begin{equation} 
  \arg (\langle k | D_{\alpha} | j \rangle) - \arg (\langle k | D_{\beta} | 
     j \rangle) = \pm \frac{\pi}{2}. \label{pi2} 
\end{equation} 
Thus, any two numbers in the set $\{\langle k | D_{\alpha} | j \rangle\}$ are 
orthogonal in the complex plane.  Since the complex plane is 
{\em two--dimensional}, at most {\em two} of the $\langle k | D_{\alpha} | j 
\rangle$ are nonzero.  If there are two, they have the promised phase 
difference of $\frac{\pi}{2}$.  Had we assumed that the histories exhibited 
{\em medium} decoherence, we would have used the decoherence condition 
(\ref{stdec}) and in (\ref{orth}) we would not have taken the real part; then 
at most {\em one} member of the set $\{\langle k | D_{\alpha} | j \rangle\}$ 
would be nonvanishing. \hfill $\Box$ \vspace{.2in}

If the initial state of the system is pure, then the theorem is still valid if 
we replace $| j \rangle \langle j |$ with $\rho$.  Thus, if the initial state 
is pure and the set $\{C_{\alpha}\}$ exhibits weak (medium) decoherence, then 
at most two histories connect (one history connects) the initial state to any 
fine-grained event with nonzero probability.  

An immediate consequence of this theorem is interesting enough to be a theorem 
of its own. \vspace{.2in} \\
{\bf Theorem 2.}  If $t_{j} < t_{k} < t_{l}$ and a nonzero-probability history 
leads to a fine-grained event at $t_{j}$ which does not occur at $t_{k}$ but 
occurs again at $t_{l}$, then no set of histories containing these events can 
decohere. \vspace{.1in} \\
{\bf Proof.}  At least two nonzero-probability histories must connect the event
at $t_{j}$ to its twin at $t_{l}$.  Further, the product of the amplitude for 
one history and the complex conjugate of that for the other is real (and 
positive), because the factors linking $t_{j}$ to $t_{k}$ are the complex 
conjugates of those linking $t_{k}$ to $t_{l}$.  Thus condition (\ref{orth}) of
Theorem~1 cannot be satisfied. \hfill $\Box$ 
\vspace{.2in} 

With Theorem~1 in hand we can immediately describe all possible decohering 
sets of histories of a two-level system (spin $\frac{1}{2}$) with a pure 
initial state.  All sets of histories with one event after the initial state 
exhibit (medium) decoherence automatically; thus we begin by considering 
two-event sets.  We assume the system is initially polarized in the direction 
{\boldmath $\vec{\imath}$}, polarized parallel or antiparallel to {\boldmath 
$\vec{n}$} at $t_{1}$, and parallel or antiparallel to {\boldmath $\vec{f}$} at
$t_{2}$.  Writing the corresponding projection operators in the standard way 
using Pauli matrices, one discovers that the weak decoherence condition 
(\ref{decohere1}) becomes
\begin{equation} 
  ( \mbox{\boldmath $\vec{\imath}$} \times \mbox{\boldmath $\vec{n}$} ) \cdot
  ( \mbox{\boldmath $\vec{n}$} \times \mbox{\boldmath $\vec{f}$} ) = 0. 
\label{geocond}  
\end{equation} 
(This result is not new \cite{om2}.)  Only these sets of two-event histories 
weakly decohere.  Further, every decohering set of histories with three or more
events is a congruent extension of a two-event set; if it were not, the number 
of nonzero-probability histories would be at least five, so at least three 
would lead from the initial state to one of the two final events, which 
Theorem~1 does not allow.  

If we were to impose {\em medium} decoherence instead, the allowed sets of 
histories would simplify considerably.  Theorem~1 allows at most one 
nonzero-probability history to lead from the initial to each of the final 
states; thus the total number of nonzero-probability histories would be at most
two.  Any set of histories which is not a congruent extension of a one-event 
set will have at least three nonzero-probability histories; thus it would not 
decohere.  Therefore the only sets of fine-grained histories of a two-level 
system which exhibit medium decoherence are one-event sets and their congruent 
extensions.  In Griffiths' language, we have shown that weakly decohering sets 
of histories corresponding to the graph in Figure~\ref{twolev}(a) exist, but 
the only sets exhibiting medium decoherence are represented by graphs like the 
one in Figure~\ref{twolev}(b), a congruent extension of a one-event set.

We call an event in a trajectory graph {\em connected} if its node leads back 
to the initial state through at least one path (if at least one history leading
to the event from the initial state occurs with nonzero probability).  We call 
it {\em singly} connected if exactly one path leads back to the initial state, 
{\em doubly} connected if two paths lead back to the initial state.  In these 
terms, Theorem~1 demands that every fine-grained event in a decohering set of 
histories be at most doubly connected (or singly connected if the set exhibits 
medium decoherence).  An event is unconnected iff it has no overlap with the 
connected events at the previous time; thus the unconnected events at any time 
lie in the span of the unconnected events at the previous time.  Therefore the 
number of {\em connected} events is a nondecreasing function of time.  
\vspace{.2in} \\
{\bf Theorem 3.}  In every transition between times in a decohering set of 
     histories represented by a trajectory graph, either
\begin{enumerate}
\item the connected events before and after the transition are identical;
\item the number of connected events increases by at least one;
\item the number of doubly connected events increases by at least {\em two}; 
      or
\item both 2 and 3 occur.
\end{enumerate}
\vspace{.1in} 
{\bf Proof.}  All we need to prove is that if 1 and 2 do not occur, then 3 must
occur.  Thus, suppose the connected events before and after the transition from
time $t_{j}$ to time $t_{j+1}$ are not identical, yet the number of connected 
events does not increase.  Then at least one event at $t_{j+1}$ must be 
connected to two events at $t_{j}$, as shown in Figure~\ref{unfor}(a).  
However, that one event at $t_{j+1}$ cannot be the {\em only} one linked to two
events at $t_{j}$; since the first event at $t_{j}$ is connected to only the 
first event at $t_{j+1}$, the two differ at most by a phase, and because the 
first and second events at $t_{j}$ are orthogonal, the first event at $t_{j+1}$
and the second event at $t_{j}$ must also be orthogonal.  Thus at least {\em 
two} events at $t_{j+1}$ must be connected to two events (each) at $t_{j}$, as 
shown in Figure~\ref{unfor}(b).  None of the doubly connected events from 
$t_{j}$ can be involved in this part of the transition (since that would make 
one of the events at $t_{j+1}$ at least triply connected); therefore, the 
number of doubly connected events increases by at least two. \hfill $\Box$ 
\vspace{.2in}

In a set of histories represented by a trajectory graph, the system's behavior 
is specified at only a finite number of times.  We might have hoped to better 
approximate continuous time evolution by inserting additional sets of events 
between those already in the graph.  However, as the next theorem shows, 
the possibilities for this are very limited. \vspace{.2in} \\
{\bf Theorem 4.}  Suppose that between times $t_{j}$ and $t_{j+1}$ in a 
     decohering set of histories represented by a trajectory graph, exactly
     one step of change occurs:  either the number of connected events 
     increases by {\em exactly} one or the number of doubly connected events 
     increases by {\em exactly} two (but not both).  Then if an additional set 
     of events is inserted between $t_{j}$ and $t_{j+1}$ while maintaining 
     decoherence, it must be identical to either the set at $t_{j}$ or the
     set at $t_{j+1}$.  \vspace{.1in} \\
{\bf Proof.} Suppose that the new set is identical to neither the set before 
nor the set after.  Then in the transition from $t_{j}$ to $t_{j+1}$ at least 
two steps of change must occur (one for the transition from $t_{j}$ to the 
intermediate time, one for the transition from the intermediate time to 
$t_{j+1}$). \hfill $\Box$ \vspace{.2in}

The histories are restricted even more drastically if only {\em finitely} many 
events occur with nonzero probability (so only that many events are connected).
\vspace{.2in} \\
{\bf Theorem 5.}  Consider a trajectory graph representing a set of decohering 
     histories with a finite number $n$ of connected events at a particular 
     time.  Excluding congruent extensions, the number of transitions prior to 
     that time is at most \mbox{$n + [\frac{n}{2}] - 2$}, where [ ] denotes the
     greatest integer part. 
\vspace{.1in} \\
{\bf Proof.}  Suppose that the given set of decohering histories contains no 
congruent extensions.  The number of connected events at time $t_{1}$ is 
therefore at least two, so the number of transitions that increase the number 
of connected events is at most \mbox{$n - 2$}.  The number of transitions that 
increase the number of doubly connected events is at most $[\frac{n}{2}]$.  
Thus the total number of transitions is at most \mbox{$n + [\frac{n}{2}] - 2$}.
\hfill $\Box$ \vspace{.2in} \\
This bound is the strongest possible, because for every $n$ there is a set of 
decohering histories in an $n$-dimensional space with this maximum number of
noncongruent steps (the $n = 5$ case is illustrated in Figure~\ref{maxnon}).
The consequences of this theorem are avoided only if the number of connected 
events is infinite right at the start, so that infinitely many events occur 
with nonzero probability at each time.

Comparison with Figure~\ref{twolev} shows that in each of the last two 
transitions in Figure~\ref{maxnon} the system can be decomposed into two 
subspaces, in one of which the transition is to congruent events while in the 
other the transition is that of a two-level system.  In fact, a large class of 
transitions is of this general type, as we show with our final theorem.
\vspace{.2in} \\
{\bf Theorem 6.}  In every transition in which the number of connected events 
     is finite and does not increase, the matrix describing the transition 
     between the {\em connected} events is block-diagonal (to within 
     rearrangement of the rows and columns), and each block is either \mbox{$2 
     \times 2$} or \mbox{$1 \times 1$}. \vspace{.1in} \\ 
{\bf Proof.}  Let the transition from $t_{j}$ to $t_{j+1}$ leave the number of 
connected events $n$ unchanged.  Since the span of the connected events at 
$t_{j}$ lies in the span of the connected events at $t_{j+1}$ and both have 
dimension $n$, the two subspaces are the same; so they have the same orthogonal
complement.  Thus each (un)connected event at $t_{j+1}$ overlaps {\em only} the
(un)connected events at $t_{j}$.  Since each connected event at $t_{j+1}$ is at
most doubly connected, each is linked to either one or two connected events at 
$t_{j}$ and no others; thus the matrix describing the entire transition has at 
most two nonzero entries in each column representing a connected event at 
$t_{j+1}$.  If a column has only one nonzero entry, then unitarity guarantees 
that the entry is also the only nonzero entry in its row; this yields all of 
the \mbox{$1 \times 1$} blocks.  If a column has two nonzero entries, then the 
orthogonality of different rows and columns demands that the entries in the 
same two rows of {\em one} and {\em only} one other column are also nonzero.  
Those two columns together form a \mbox{$2 \times 2$} block; all other entries 
in their rows and columns are zero. \hfill $\Box$ \vspace{.2in} \\
This theorem reduces the allowed transitions to an extremely simple form; its 
restrictions are avoided only if the number of connected events (the number of 
events that occur with nonzero probability) increases continually over time.

These results suggest that the decoherence conditions strongly favor histories 
dominated overwhelmingly by congruent extensions.  It is not surprising that 
decoherence selects out the histories that conform with the system's unitary 
evolution, but the extent to which they are preferred is remarkable.  For 
example, only congruent events can occur between congruent events (Theorem~2), 
and if continuous classical evolution is to be approached by inserting events 
at more and more times, almost all insertions must be congruent extensions
(Theorems 2, 4, and 5).  Probabilities that are periodic in time and are for a 
finite number of events at some time must be for congruent events and are 
therefore constant in time.  (The number of connected events can not decrease, 
so if it is periodic it must be constant and Theorem 6 applies.)

\begin{thebibliography}{9}

\bibitem{SFI} M. Gell-Mann and J. B. Hartle, in {\it Complexity, Entropy and 
              the Physics of Information}, ed. W. Zurek, Santa Fe Institute 
              Studies in the Science of Complexity, Vol. VIII (Addison-Wesley, 
              Reading, 1990).

\bibitem{misner} J. B. Hartle, in {\it Directions in General Relativity:  
                 Proceedings of the 1993 International Symposium}, ed. B. L. 
                 Hu, M. P. Ryan, and C. V. Vishveshwara, Vol. 1 (Cambridge 
                 University Press, Cambridge, 1993).

\bibitem{D47} M. Gell-Mann and J. B. Hartle, Phys. Rev. D {\bf 47}, 3345 
              (1993).

\bibitem{consis} R. B. Griffiths, J. Stat. Phys. {\bf 36}, 219 (1984).

\bibitem{graph} R. B. Griffiths, Phys. Rev. Lett. {\bf 70}, 2201 (1993).

\bibitem{om1} R. Omn\`{e}s, J. Stat. Phys. {\bf 53}, 893, 933, 957 (1988).

\bibitem{om2} R. Omn\`{e}s, Rev. Mod. Phys. {\bf 64}, 339 (1992). \ See also 
              the substantial list of references therein.

\end{thebibliography}
\end{document}